\documentclass[sigconf]{acmart}

\AtBeginDocument{%
  }

\setcopyright{acmlicensed}
\copyrightyear{2018}
\acmYear{2018}
\acmDOI{XXXXXXX.XXXXXXX}

\acmConference[Conference acronym 'XX]{Make sure to enter the correct
  conference title from your rights confirmation emai}{June 03--05,
  2018}{Woodstock, NY}
\acmISBN{978-1-4503-XXXX-X/18/06}

\begin{document}

\title{Generate the browsing process for short-video recommendation}

\author{Chao Feng}
\authornote{Authors contributed equally to this research.}
\email{chaofeng@mail.ustc.edu.com}
\affiliation{%
  \institution{Kuaishou Technology}
  \city{Beijing}
  \country{China}
}

\author{Yanze Zhang}
\authornotemark[1] 
\email{yanze3@gmail.com}
\affiliation{%
  \institution{Kuaishou Technology}
  \city{Beijing}
  \country{China}
}

\author{Chenghao Zhang}
\authornotemark[1] 
\authornote{Work done while interning at Kuaishou Technology.}
\email{harryzhang719@gmail.com}
\affiliation{%
  \institution{Kuaishou Technology}
  \city{Beijing}
  \country{China}
}

\renewcommand{\shortauthors}{Trovato et al.}

\begin{abstract}
This paper proposes a generative method to dynamically simulate users' short video watching journey for watch time prediction in short video recommendation. Unlike existing methods that rely on multimodal features for video content understanding, our method simulates users' sustained interest in watching short videos by learning collaborative information, using interest changes from existing positive and negative feedback videos and user interaction behaviors to implicitly model users' video watching journey. By segmenting videos based on duration and adopting a Transformer-like architecture, our method can capture sequential dependencies between segments while mitigating duration bias. Extensive experiments on industrial-scale and public datasets demonstrate that our method achieves state-of-the-art performance on watch time prediction tasks. The method has been deployed on Kuaishou Lite, achieving a significant improvement of +0.13\% in APP duration, and reaching an XAUC of 83\% for single video watch time prediction on industrial-scale streaming training sets, far exceeding other methods. The proposed method provides a scalable and effective solution for video recommendation through segment-level modeling and user engagement feedback.
\end{abstract}



\received{20 February 2007}
\received[revised]{12 March 2009}
\received[accepted]{5 June 2009}

\maketitle

\section{Introduction}
With the rapid growth of short video platforms such as TikTok and Kuaishou, video recommendation systems play an increasingly important role in enhancing user experience and retention rates. Unlike traditional click-through rate (CTR) metrics, watch time has become a key indicator of user interest and satisfaction in short video scenarios, where videos are automatically played in a feed-like manner. Accurately predicting user watch time is crucial for optimizing video recommendation rankings and improving platform revenue. Furthermore, the ordinal relationships in watch time prediction are crucial as they reflect users' relative preferences for different videos \citeauthor{Lin_2022}.

However, watch time prediction faces several unique challenges. First, watch time exhibits a highly skewed long-tail distribution, where a few popular videos account for the majority of exposures while most videos have insufficient exposure, leading to extremely imbalanced data. Second, watch time depends not only on the match between user interest and video quality, but is also influenced by the video duration itself. Users tend to spend more time watching longer videos, and this duration bias may mislead models to recommend longer videos while ignoring users' true interest preferences.

Moreover, user watching behavior is a generative behavior with significant sequential dependencies, meaning that watched segments affect the willingness to continue watching subsequent segments. Modeling this sequential dependency is crucial for accurately predicting watch time, as it captures the dynamic nature of user engagement. Traditional methods often ignore this conditional dependency between segments, leading to poor performance. Most existing approaches are merely regressive or use tree-based discriminative direct prediction of watch time, which may be affected by error accumulation in autoregressive models. This cumulative error problem significantly reduces the quality of watch time prediction. In short video user watching behavior, users do not "decide how long to watch" in a single click, but continuously make micro-decisions of "whether to continue" at the boundary of each segment; watch time is merely the natural "generated" result of this series of decisions along the time axis.

Based on this observation, we reshape "predicting how long to watch" from discriminative point estimation to a "generating watching process" modeling task: through user historical watching behavior and positive/negative feedback information from other users on video segments, the model generates segment-wise conditional probabilities of continuing to watch along the time axis, forming an interpretable and calibratable survival/retention curve, where watch time is merely the discrete summation/integration result of this curve. This formulation naturally preserves sequential dependency information, structurally weakens duration-induced bias, and allows parallel inference to meet industrial constraints. Furthermore, we rely solely on interaction behavior modeling to predict user segment-level interest for video watch time prediction, decomposing video ID embeddings into segment representations, combining user-video cross attention with positional bias/monotonic priors, achieving segment-level modeling and duration reconstruction without introducing heavy multimodal chains. Extensive experiments and large-scale online A/B tests demonstrate that this generative paradigm achieves stable and significant gains in watch time prediction. The method has been deployed on Kuaishou Lite, achieving a significant improvement of +0.13\% in APP duration, and reaching an XAUC of 83\% for watch time prediction on industrial-scale streaming training sets, far exceeding other methods. The proposed method provides a scalable and effective solution for video recommendation through segment-level modeling and user engagement feedback.

\section{Related Works}
\subsection{Watch Time Prediction}
Watch time prediction has gained significant attention in the video recommendation domain due to its importance in measuring user engagement and satisfaction. One of the pioneering works in this area is the YouTube DNN model proposed by \citeauthor{45530} (2016). This model treats watch time prediction as a weighted logistic regression problem, where the positive samples (watched videos) are weighted by their watch time. However, this approach assumes a linear relationship between watch time and user interest, which may not hold in practice.

To address the limitations of the YouTube DNN model, \citeauthor{zhan2022deconfoundingdurationbiaswatchtime} (2022) introduced the Duration-Deconfounded Quantile-based (D2Q) method. D2Q aims to mitigate the duration bias in watch time prediction by grouping videos based on their duration and predicting watch time quantiles within each group. By focusing on the relative ordering of watch time rather than the absolute values, D2Q reduces the influence of video length on the prediction. However, D2Q does not explicitly model the sequential dependency of user watching behavior.

Other approaches to watch time prediction include the use of survival analysis (Yu et al., 2019) and hierarchical recurrent neural networks (Wu et al., 2018). These methods attempt to capture the temporal dynamics of user engagement and predict the likelihood of a user continuing to watch a video. However, they do not directly address the issues of duration bias or sequential dependency.
\subsection{Ordinal Regression}
Ordinal regression is a technique that aims to predict ordinal labels, where the relative order of the labels is important. In the context of watch time prediction, ordinal regression can be used to model the ordinal nature of user engagement levels. The Tree-based Progressive regression Model (TPM) proposed by \citeauthor{lin2023treebasedprogressiveregression} is an example of applying ordinal regression to watch time prediction. TPM divides videos into segments and models the conditional probability of a user watching each segment given the previously watched segments. By considering the sequential dependency of user behavior, TPM captures the dynamic patterns of user engagement.

Other ordinal regression approaches, such as the Ordinal Regression (OR) model\cite{7780901} and the Ordinal Regression with Multiple Output CNN (OR-CNN) \cite{fu2018deepordinalregressionnetwork}, have been applied to various tasks, including age estimation and image ranking. These methods demonstrate the effectiveness of ordinal regression in handling ordered labels and capturing the relative preferences of users. However, they have not been extensively explored in the context of watch time prediction in video recommendation.
\subsection{Interest Signal}
A concurrent work focuses on segment-level dynamic interest modeling for short videos\citeauthor{he2025shortvideosegmentleveluser}: it adopts a "hybrid representation—multimodal user-video encoder—segment interest decoder" backbone, fusing user ID/history with video visual content, and outputs segment interest scores through user-video cross attention. While this work shares some conceptual similarities with ours, we are more specifically focused on watch time prediction, a specialized subdomain in recommendation systems, though it also provides a broader perspective for our exploration. Beyond the different exploration directions, we have two key differences: (1) If we view our method as interest signal modeling, their approach directly predicts "segment interest" and uses it as weights for downstream services, without explicitly modeling the complete continuation viewing process; our work generates segment-wise continuation viewing probabilities (interest signals)/survival curves, with watch time as their natural aggregation. (2) Their approach adopts multimodal features (such as CLIP visual features), making training and online service chains heavier; while our work relies solely on collaborative signals (ID and behavior), achieving stable gains and low-latency deployment without requiring multimodal features. Overall, the former emphasizes multimodal fusion to characterize "interest intensity," while our work uses "process curves" to characterize the viewing generation mechanism, emphasizing pure collaboration, lightweight chains, and deployability.


\begin{figure*}[htbp]
  \centering
  \includegraphics[width=\textwidth]{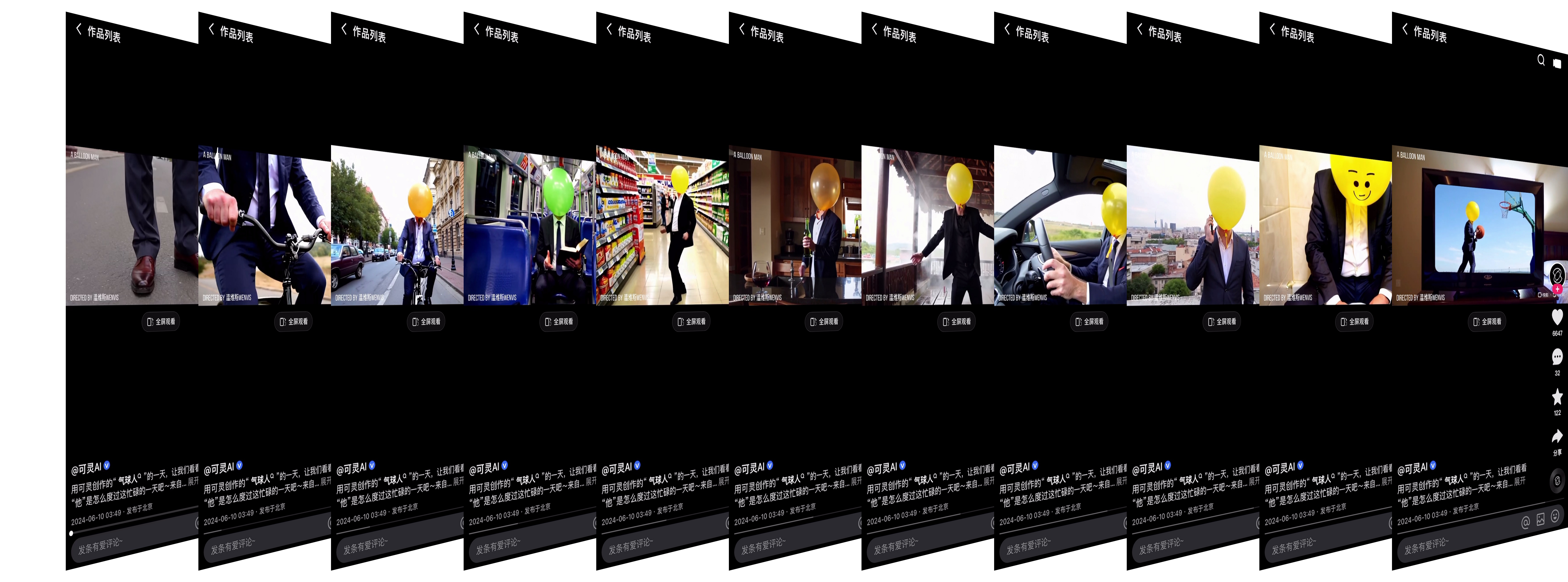}
  \caption{The video is segmented into $M$ clips $m_0, m_1, \ldots, m_{M-1}$ as illustrated, and the user has watched up to the $i$-th clip, the probability of the user continuing to watch the $(i{+}1)$-th clip can be modeled as a conditional probability:$\operatorname{Prob}\left(m_{i+1} \mid m_0, m_1, \ldots, m_i\right)$
}
  \label{fig:your_label}
\end{figure*}

\begin{figure*}[htbp]
    \centering
    \includegraphics[width=\linewidth]{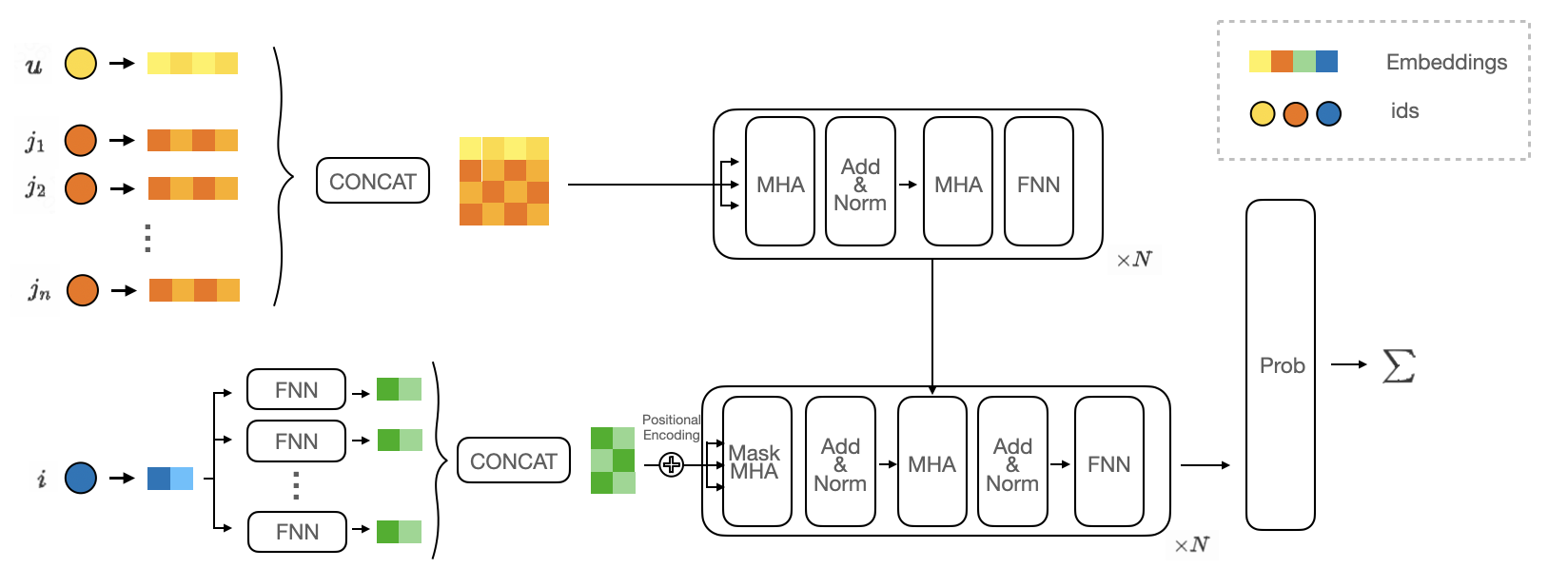}
    \caption{Architecture of our Generative Model of User Viewing Process. The model predicts user watch time by modeling how users decide to continue watching a video segment by segment. At the top, user ID and behavior sequence are embedded and encoded to capture user intent. At the bottom, the target video is split into segments, each represented through a feed-forward network and combined with positional information. These segment representations are used to predict the probability $p_i$ of watching each segment $m_i$. The final watch time is computed as $\sum_{i=0}^{M-1} p_i \cdot d_i$, where $d_i$ is the duration of segment $m_i$.}
    \label{fig:enter-label}
\end{figure*}

\section{Method}
\subsection{Problem Definition}
On short video platforms, users continuously watch videos lasting from a few seconds to several minutes. Videos are displayed one by one, and users swipe to view the next video recommended by the system. Formally, let $U = \{u_1, u_2, ...\}$ denote the set of users and $V = \{v_1, v_2, ...\}$ denote the set of videos. 

\textbf{Video segments} are defined through uniform time interval partitioning: each video $v \in V$ is divided into $M$ segments, $S_v = (S_v^1, S_v^2, ..., S_v^M)$, where each segment $S_v^i$ has a duration of $t$. The watched segments are determined based on the watch time of each video, and we assume users start watching videos from the beginning, as this is common behavior in video feeds. Given a user $u$ and a target video $v$, our goal is to infer the watching probability sequence of video segments $S_v$: $\vec{p} = (p_1, p_2, ..., p_M)$, where $p_i$ is the watching probability of segment $S_v^i$.

\subsection{Unified Perspective of Probability and Duration}
We unify segment-level "continuation viewing" probabilities, survival curves, and expected duration under the same perspective: let $p_i = P(\text{continue watching segment } i | \text{watched up to segment } i-1)$, then the sequence $\{p_i\}$ formed along the time axis constitutes a discrete survival curve; the watch time can be written as $T = \sum_{i=0}^{M-1} p_i \cdot d_i$, where $d_i$ is the duration of the $i$-th segment. When the segment width approaches finer granularity, the above expression approximates the continuous-time integral expression. To align with real viewing patterns, we introduce positional bias in the model and impose monotonic regularization constraints on the prior that $p_i$ decreases with segment position, while appropriately calibrating the output probabilities so they can be used both for duration reconstruction and as robust weights for ranking and decision-making.

\subsection{Model Overview}
As shown in Figure 2, the collaborative user-video representation learner constructs user representations based on user IDs and their historical behaviors at first; simultaneously, the target video's ID and segment position indices are input to a segment-aware MLP  to obtain chronologically ordered segment-level representations $V_1...V_M$. Subsequently, the segment-aware sequence encoder takes user historical sequences and target video segment sequences as input, obtaining segment representations that fuse user preferences through user-video cross attention and lightweight self-attention. Finally, the duration prediction decoder outputs the continuation viewing probability for each segment under positional bias and monotonic prior constraints, and performs weighted aggregation over the entire probability curve to obtain the expected duration.

The design of this structure is based on three considerations: First, segment decomposition transforms "duration-induced bias" into a structural problem, enabling the model to directly express "continuation viewing" conditional probabilities at the segment level, fundamentally weakening the preference for longer videos; Second, cross attention aligns user history with target segments without introducing heavy multimodal components, capturing short-term dependencies that affect continuation viewing; Finally, monotonic and positional priors at the decoder end provide necessary constraints for sparse data, combined with parallel computation to bring stable, low-latency online availability. Overall, the model outputs an interpretable, calibratable, and reusable survival/retention probability curve that can directly serve downstream tasks such as duration aggregation, skip prediction, and ranking weighting.

\subsection{Collaborative User-Video Representation Learning}
\subsubsection{User Representation Learning}
User representations are learned through collaborative information, including user ID embeddings and historical behavior sequence modeling. User ID embeddings capture users' static preference features:

$$U_{id} = \text{Embedding}(\text{user\_id})$$

User historical behavior sequences are encoded through self-attention mechanisms to learn users' dynamic interest patterns:

$$H = [h_1, h_2, ..., h_L]$$
$$\text{ where } h_i = \text{Embedding}(\text{video}_i) + \text{Embedding}(\text{behavior}_i)$$

$$U_{seq} = \text{MultiHeadAttention}(H, H, H)$$

The final user representation integrates static and dynamic features through a gated fusion mechanism:

$$U = \sigma(W_g[U_{id}; U_{seq}]) \odot U_{id} + (1 - \sigma(W_g[U_{id}; U_{seq}])) \odot U_{seq}$$

\subsubsection{Video Segment Representation Decomposition}
We decompose the overall video embedding into $M$ segment-level representations through segment-aware MLP networks:

$$\begin{aligned}
\mathbf{V}_i^{(0)} &= \underbrace{\mathbf{W}_2\,\sigma\!\left(\mathbf{W}_1\,[\,\mathbf{v}_{\text{video}}\,;\,\mathbf{e}_i\,] + \mathbf{b}_1\right) + \mathbf{b}_2}_{\text{Segment-specific MLP}_i}\; +\; \underbrace{\text{PE}(i)}_{\text{Positional encoding}},\\
\mathbf{e}_i &= \text{Emb}(i),\; \sigma=\text{GELU},\;\; \mathbf{W}_1\in\mathbb{R}^{h\times(d_v+d_p)},\; \mathbf{W}_2\in\mathbb{R}^{d\times h}.
\end{aligned}$$

where $\text{PE}(i)$ is positional encoding, and $\text{MLP}_i$ is the $i$-th segment-specific multi-layer perceptron. This design enables the model to learn unique features of segments at different positions while avoiding expensive multimodal feature extraction.

\subsection{Segment-Aware Sequence Encoder}
The segment-aware sequence encoder is responsible for modeling the interaction relationships between users and video segments, adopting a Transformer-like architecture to process sequence information.

\subsubsection{User-Video Cross Attention}
The core of the encoder is the user-video cross attention mechanism. Unlike traditional Transformers, we separately handle the interactions between user sequences and video segment sequences. In the $l$-th layer:

$$Q_V^{(l)} = V^{(l-1)}W_Q^V, \quad K_U^{(l)} = U^{(l-1)}W_K^U, \quad V_U^{(l)} = U^{(l-1)}W_V^U$$

$$Q_U^{(l)} = U^{(l-1)}W_Q^U, \quad K_V^{(l)} = V^{(l-1)}W_K^V, \quad V_V^{(l)} = V^{(l-1)}W_V^V$$

Cross attention weight computation:

$$A_{UV}^{(l)} = \text{Softmax}(Q_V^{(l)}(K_U^{(l)})^T / \sqrt{d_k})$$

$$A_{VU}^{(l)} = \text{Softmax}(Q_U^{(l)}(K_V^{(l)})^T / \sqrt{d_k})$$

Self attention weight computation:

$$A_{VV}^{(l)} = \text{Softmax}(Q_V^{(l)}(K_V^{(l)})^T / \sqrt{d_k})$$

$$A_{UU}^{(l)} = \text{Softmax}(Q_U^{(l)}(K_U^{(l)})^T / \sqrt{d_k})$$

\subsubsection{Representation Update and Fusion}
User and video representations are updated through the combination of cross attention and self attention:

$$V'^{(l)} = A_{UV}^{(l)}V_U^{(l)} + A_{VV}^{(l)}V_V^{(l)}$$

$$U'^{(l)} = A_{VU}^{(l)}V_V^{(l)} + A_{UU}^{(l)}V_U^{(l)}$$

Through residual connections and layer normalization, we obtain the input for the next layer:

$$V^{(l)} = \text{LayerNorm}(V'^{(l)} + V^{(l-1)})$$

$$U^{(l)} = \text{LayerNorm}(U'^{(l)} + U^{(l-1)})$$

After $L$ layers of encoding, we obtain segment representations that fuse user preferences.

\subsection{Duration Prediction Decoder}
The duration prediction decoder generates the viewing probability for each segment based on encoder outputs through sequential conditional probability modeling.
\subsubsection{Segment Interest Modeling}
The encoder outputs are mapped to segment interest scores through MLP:

$$h_i = \text{MLP}(V_i^{(L)}), \quad i \in \{0, 1, ..., M-1\}$$

To model the interaction relationships between segments, we adopt bilinear fusion:

$$\begin{aligned}
z_i &= \mathbf{w}_1^\top \mathbf{h}_i + \sum_{j\neq i}\mathbf{h}_i^\top\mathbf{A}\,\mathbf{h}_j + \sum_{k=1}^{r}\langle \mathbf{p}_k,\mathbf{h}_i\rangle\,\langle \mathbf{q}_k,\mathbf{h}_i\rangle + b_f,\\
\mathbf{A} &= \sum_{k=1}^{r}\lambda_k\,\mathbf{p}_k\mathbf{q}_k^\top,\;\; r\ll d,\}.
\end{aligned}$$

\subsubsection{Temporal Position Modeling}
Considering the temporal decay characteristics of user viewing behavior, we introduce positional bias modeling:

$$s_i = z_i + w_p \cdot \phi(i) + b_p$$

where $\phi(i)$ is the encoding function for position $i$, and $w_p$ and $b_p$ are learnable positional bias parameters.

\subsubsection{Sequential Conditional Probability}

The final viewing probability is computed through sequential conditional modeling, considering the viewing history of previous segments:

$$P(m_i | m_0, ..., m_{i-1}) = \sigma(s_i + \gamma \cdot \sum_{j=0}^{i-1} P(m_j) \cdot \beta_j)$$

where $\gamma$ and $\beta_j$ are learnable parameters used to model the influence of previous viewing behavior on the current segment.

$$\begin{aligned}
q_i &= \sigma\!\left(\frac{s_i}{\tau}\right),\;\; \text{(temperature } \tau \text{ smoothing)}\\
\underbrace{P(m_i\mid m_{<i})}_{p_i} &= \prod_{t=0}^{i} q_t,\;\; \underbrace{S_i}_{\text{survival probability}}=\prod_{t=0}^{i-1}(1-q_t),\\
\text{or}\; p_i &= \sigma\!\big(s_i + \gamma\sum_{j=0}^{i-1}k(i,j)\,p_j\big),\; k(i,j)=\exp\!\big(-\alpha(i-j)\big).
\end{aligned}$$

\subsubsection{Watch Time Aggregation}

The user's total watch time is obtained through weighted summation of segment probabilities:

$$\begin{aligned}
\mathbb{E}[T] &= \sum_{i=0}^{M-1} S_i\,\Delta t_i,\;\; S_i=\prod_{t=0}^{i-1}(1-q_t),\; \Delta t_i=d_i,\\
\lim_{\max d_i\to 0}\; \mathbb{E}[T] &= \int_0^{t_{\max}} S(t)\,\mathrm{d}t,\;\; S(t)=\exp\!\left(-\int_0^t \lambda(u)\,\mathrm{d}u\right).
\end{aligned}$$

where $d_i$ represents the duration of the $i$-th segment.

This design enables the model to: (1) compute all segment probabilities in parallel, avoiding cumulative errors of autoregressive models; (2) explicitly model sequential dependencies between segments; (3) implicitly capture video content features through collaborative information.

\subsubsection{Loss Function}

During inference, the model can compute all segment viewing probabilities in parallel, avoiding the cumulative error problem of autoregressive models. Given a user and target video, the model outputs segment probability sequences and total watch time predictions.

The model training process involves optimizing parameters of the Transformer-like encoder-decoder architecture to minimize a composite loss function. The loss function contains three components: cross-entropy loss for classification, Huber loss for watch time recovery, and ordinal loss for enforcing monotonicity of segment-level predictions.

\textbf{Cross-entropy Loss}: For each video segment $m_i$, the model predicts the probability $p_i$ that a user continues watching that segment given previously watched segments. This is formulated as a binary classification problem. We use cross-entropy loss to measure the difference between predicted probabilities and true labels.
$$\begin{aligned}
\mathcal{L}_{\text{seq}} &= -\sum_{(u,v)}\sum_{i=1}^{M-1} w_i\,\log p_i,\;\; w_i=\omega_0\,\exp(-\rho i)
\end{aligned}$$
\textbf{Huber Loss}: To enforce accurate recovery of watch time from segment-level predictions, we adopt Huber loss. Compared to mean squared error loss, Huber loss is less sensitive to outliers.
$$\begin{aligned}
\mathcal{L}_{\text{Huber}}(T,\hat T) &= \begin{cases}
\frac{1}{2}(T-\hat T)^2,& |T-\hat T|\le \delta,\\
\delta\big(|T-\hat T|-\frac{1}{2}\delta\big),& \text{otherwise},\end{cases}
\end{aligned}$$
\textbf{Ordinal Loss}: Since segments follow temporal order, predicted probabilities should monotonically decrease as segment indices increase. This loss penalizes violations of monotonicity constraints by considering differences between adjacent segment probabilities.
$$\begin{aligned}
\mathcal{L}_{\text{ord}} &= \sum_{i=0}^{M-2} \max\big\{0,\, p_{i+1}-p_i+\varepsilon\big\} + \mu\sum_{i=1}^{M-1}|p_i-p_{i-1}|
\end{aligned}$$
The final training objective is a weighted combination:

$$\begin{aligned}
\mathcal{L} &= \lambda_1\mathcal{L}_{\text{seq}}+\lambda_2\mathcal{L}_{\text{Huber}}+\lambda_3\mathcal{L}_{\text{ord}}.
\end{aligned}$$\\

Through this generative modeling framework, our method can effectively capture the dynamic characteristics of user viewing behavior while avoiding the complexity of traditional multimodal methods. The model relies solely on collaborative information to achieve fine-grained segment-level modeling, providing a scalable and efficient solution for watch time prediction.

\section{Experiments}

In this section, we evaluate our framework on benchmark recommendation datasets, including two public recommendation datasets (WeChat and KuaiRec) and a real-world industrial dataset (Industrial). Additionally, we conduct ablation studies to validate the design choices of the proposed framework, training strategies, and hyperparameter selections.

\subsection{Experimental Setup}
For fair comparison, we adopt offline experiments on industrial datasets and two public datasets to ensure reproducibility. The large-scale industrial dataset (called Industrial) is collected from a real-world streaming short video platform. This dataset contains approximately 6 billion impressions and over 50 million users per day. We accumulate 14 days of user interaction logs and subsequently use the next day's information for evaluation. Furthermore, to improve reproducibility, we adopt public benchmark datasets KuaiRec and WeChat. The KuaiRec dataset contains a total of 7,176 users, 10,728 distinct items, and a substantial 12,530,806 impressions, while the evaluation dataset contains a collection of 1,411 users, 3,327 items, and 4,676,570 impressions. Similarly, the WeChat dataset contains user IDs, video IDs, and user interaction behaviors.

\textbf{Evaluation Metrics}: We follow previous work and adopt the following metrics to quantitatively measure watch time prediction:
\\\\
MAE (Mean Absolute Error): Measures the average absolute error between watch time predictions and ground truth values.\\
$$
\mathrm{MAE} = \frac{1}{N} \sum_{i=1}^{N} \left| \hat{y}_i - y_i \right|
$$
XAUC: Measured by randomly sampling instance pairs and checking whether their relative order is consistent with the ground truth.
$$
\mathrm{XAUC} = \frac{1}{|\mathcal{P}|} \sum_{(i,j)\in \mathcal{P}} \mathbf{1}\!\left\{ \operatorname{sign}(\hat{y}_i - \hat{y}_j) = \operatorname{sign}(y_i - y_j) \right\}
$$
\\

\textbf{Baseline Methods}: Here, we compare our method with previous approaches for watch time prediction tasks. The reference methods can be divided into traditional machine learning methods and neural network methods. The former includes Value Regression (VR), Weighted Logistic Regression (WLR), and Ordinal Regression (OR), while the latter includes the state-of-the-art model D2Q. Specifically, VR and WLR are continuous regression methods (denoted as Cont.), while OR and D2Q decompose the target range into discrete buckets (denoted as Disc.). Since the authors did not provide official code for watch time prediction, we implement them according to their corresponding papers, adjust the number of buckets (if applicable), and report the best results.

\subsection{Comparison with State-of-the-Art Methods}
In this section, we compare our method with current state-of-the-art watch time prediction methods and report MAE and XAUC metrics in Table 1. We can see that on the KuaiRec dataset, discrete methods outperform the optimal solution by at least 0.164 in XAUC.
\begin{table}[h]  
    \centering
    \resizebox{8cm}{!}{
    \begin{tabular}{c|cc|cc}
    \toprule
    Data &  KuaiRec &  & WeChat & \\
    Methods & MAE & XAUC & MAE & XAUC\\
    \midrule
    VR & 5.1538 & 0.5513 & 20.327 & 0.6215 \\
    WLR & 7.0120 & 0.5818 & 24.792 & 0.6520 \\
    D2Q & 4.8880 & 0.5874 & 19.934 & 0.6799 \\
    CREAD & 4.6432 & 0.6073 & 18.841 & 0.6943 \\
    ours & \bf{4.7408} & \bf{0.6237} & \bf{18.109} & \bf{0.7171} \\
    \bottomrule
    \end{tabular}
    }
    \caption{Watch time prediction results on KuaiRec, WeChat, and Industrial datasets: XAUC and MAE.}
    \label{tab:accuracy}
\end{table}

\subsection{Online A/B Testing}
In addition to offline experiments, we also conducted online A/B testing on the Kuaishou App, a streaming short video platform with over 400 million users daily. We allocated 10\% of traffic to our method and baseline D2Q respectively. We launched the experiment on the real-time system for 5 days and report the results in Table 4. As we can see, the watch time of the treatment group (referred to as the primary metric) significantly increased by 0.14\%.

\section{Discussion and Future Directions}
This work centers on the core idea of "generating viewing processes" for duration estimation: using segment-wise continuation viewing probabilities to characterize the entire viewing trajectory and aggregating them into expected duration, relying solely on collaborative signals (ID and behavior) without involving explicit video content understanding. Looking toward the future, we discuss three more comprehensible high-level directions: how to make the "process curve" more stable and accurate; how to better utilize supervision signals under sparse feedback; and how to more naturally integrate the curve with actual systems.

First, enhancement of representation and calibration. The current method obtains a discrete "continuation viewing probability curve" through equal-width segmentation, which has already brought advantages in interpretability and bias suppression. Subsequently, we can moderately improve the "resolution and reliability" of this curve without changing the overall paradigm: on one hand, making segmentation more flexible (e.g., adaptively determining segment width or number based on data distribution) to better align with real viewing rhythms; on the other hand, systematically performing probability calibration and uncertainty estimation, enabling model outputs to not only rank but also robustly convert to duration, threshold, or policy signals. Such modifications are beneficial for maintaining consistent interpretability and transferability across different scenarios and data slices.

Second, enhancement of weakly supervised learning and plug-and-play multimodality. Our training relies on weakly supervised signals such as "watched up to which segment/cumulative duration." Subsequently, while maintaining the lightweight backbone, we can introduce learning objectives and training strategies that are closer to the target: for example, combining ranking or quantile-oriented objectives to better align training with business evaluation; alleviating instability caused by sparse and noisy feedback through more robust regularization and data augmentation. For cold-start and cross-domain generalization challenges, multimodal priors (such as visual/audio/text embeddings) can be introduced in a "plug-and-play" manner through late fusion or knowledge distillation as a complement to the collaborative backbone; maintaining robustness of training and inference strategies under realistic conditions such as missing modalities and distribution drift. It should be emphasized that these extensions are optional and do not change the basic approach of "collaboration-first, parallel lightweight chains."

Third, system integration and long-term value. Process curves are naturally suited for collaboration with ranking systems: segment-level probabilities can be aggregated into video-level scores, and key statistics of the curve (such as early retention tendencies and potential churn turning points) can be incorporated as general features into existing models. Furthermore, objectives such as "duration, interaction quality, and retention" should be comprehensively considered to avoid pursuing only short-term duration at the expense of long-term satisfaction; while conducting more thorough offline and online validation across different populations and duration distributions to ensure fairness and robustness. From an engineering perspective, delay and throughput can be controlled without sacrificing effectiveness through distillation, caching, or early exit techniques, and the interpretability of curves can be used to guide policy optimization and content optimization on the creation side.

\section{Conclusion}
In this paper, we propose a novel video recommendation watch time prediction method that simulates users' video viewing journey through segment-level modeling using historical viewing behavior and adopts a Transformer-like architecture. This method can capture sequential dependencies between segments while mitigating duration bias. The proposed method achieves state-of-the-art performance on both industrial-scale and public datasets without relying on complex multimodal data for video content understanding.

The results of this method demonstrate the effectiveness of segment-level modeling and the feasibility of using user engagement feedback for implicit video modeling. This approach opens new possibilities for improving video recommendation by considering the fine-grained structure of videos and user viewing patterns.

Future research directions include combining explicit video embeddings with segment-level understanding, extending the method to other video recommendation tasks, exploring temporal evolution of user engagement, and adapting the method to various video recommendation scenarios. By continuing to research and refine segment-level modeling techniques, we can improve the performance and user experience of video recommendation systems.

\bibliography{custom}
\bibliographystyle{ACM-Reference-Format}

\end{document}